# Hyperdoped silicon photodetectors enable room-temperature computational SWIR imaging at 1550 nm

Xiaolong Liu[1,2,*], Sören Schäfer[3], Jinyuan Chen[2], Patrik Mc Kearney[3], Simon Paulus[3], Varsha Ashwin Vedaraj[2], Ville Vähänissi[1], Stefan Kontermann[3], Kenneth Crozier[2,4,5], James Bullock[2], Hele Savin[1]

1. Department of Electronics and Nanoengineering, Aalto University, 02150 Espoo, Finland
2. Department of Electrical and Electronic Engineering, University of Melbourne, Victoria 3010, Australia
3. Institute for Microtechnologies (IMtech), RheinMain University of Applied Sciences, Am Brückweg 26, 65428 Rüsselsheim, Germany
4. ARC Centre of Excellence for Transformative Meta-Optical Systems (TMOS), University of Melbourne, Victoria 3010, Australia
5. School of Physics, University of Melbourne, Victoria 3010, Australia
* Correspondence: Xiaolong Liu (xiaolong.liu@aalto.fi)

## Abstract

Silicon's bandgap inherently restricts its photodetection to wavelengths below 1100 nm, necessitating the integration of costly III-V semiconductors for short-wave infrared applications. Hyperdoping silicon beyond the solid solubility limit offers a promising "silicon-native" alternative, yet achieving practical short-wave infrared applications at room temperature remains a formidable challenge. Here, we demonstrate a high-detectivity hyperdoped silicon photodetector enabling room-temperature computational short-wave infrared imaging beyond Si bandgap wavelength at $\lambda$ = 1550 nm. By integrating an ultrafast laser heating process step to reduce the dark current while keeping high responsivity, we achieve a specific detectivity $D^*$ exceeding $10^9$ Jones for 1550 nm at room temperature working in a forward-biased, photoconductive mode. The improved detectivity, coupled with a 59.4 dB linear dynamic range and kHz-scale bandwidth, allows us to demonstrate a single-pixel imaging system that reconstructs 1550 nm scenes at 65×63 pixels without cryogenic cooling. Our devices simultaneously support visible-light imaging, offering a path toward monolithically integrated, multispectral Si-native optical sensors. These results establish ultrafast-laser hyperdoped silicon as a viable platform for low-cost, room-temperature, short-wave infrared photonics, bridging the gap between advanced materials science and practical computational imaging system.

## Introduction

Silicon is the workhorse of modern electronics and photonics, yet its bandgap confines conventional Si photodetectors to wavelengths below $\lambda$ = 1100 nm. This precludes direct detection in short-wave infrared (SWIR) that is increasingly exploited, e.g., in fiber networks[1–3], imaging/LiDAR[4–7], biochemical sensing[8–10] and free-space optical communication[11–13]. Other than silicon, accessing the SWIR with mature technologies typically requires compound semiconductors such as PbS, InGaAs and GaSb, which are toxic, expensive, difficult to monolithically integrate with Complementary Metal–Oxide–Semiconductor (CMOS) technology, and often rely on complex epitaxial growth or cooling[14–16]. These constraints hinder the development of large-area, low-cost SWIR imagers and integrated photonic–electronic systems. The fundamental challenge remains whether silicon can be engineered beyond its intrinsic bandgap limitations to achieve the sufficiently-high sub-bandgap performance required for practical, room-temperature SWIR applications without the need for heterogeneous integration.

Hyperdoping provides a conceptually attractive route for trials in extending silicon's spectral reach[17–26]. By introducing deep-level impurities at concentrations far beyond equilibrium solid solubility, hyperdoping creates intermediate-band states that enable sub-bandgap absorption and, in principle, SWIR detection using standard silicon wafers and CMOS-compatible processes. A variety of hyperdoped-Si devices have demonstrated room-temperature response beyond 1100 nm (**Supplementary Table S1**), but most work has focused near ~1000–1350 nm, where peak responsivity is located or high enough responsivity can be achieved at the expense of millisecond response times[24,25,27]. By contrast, systematic characterization at longer wavelengths remains sparse: reported responsivities are typically very low[20–22], often not even fully quantified or they are enabled by deep-cooling[23]. Moreover, practical device metrics such as noise, bandwidth and detectivity are rarely reported for longer SWIR wavelengths. This is partly because responsivities at longer wavelengths are generally much (~> 3 orders of magnitude) lower than peak responsivity at ~1000 nm, and often not comprehensively measured, either due to instrumental limitations or historical assumptions that performance at longer wavelengths is too limited to be practically relevant. As a result, across the family of hyperdoped silicon materials and devices, performance at longer SWIR bands remains under-explored and under-exploited for practical applications.

Beyond fundamental material metrics, validating the viability of hyperdoped silicon requires a demonstration of performance in practical, high-demand sensing scenarios. Computation-based single-pixel imaging (SPI), which integrates spatial light modulation with a single non-pixelated sensor, serves as an ideal benchmark. It provides a stringent system-level test of a detector's detectivity, noise profile, and bandwidth without the overhead of fabricating complex focal plane arrays. To date, however, SWIR SPI implementations have largely relied on compound semiconductors or emerging thin-film materials such as colloidal quantum dots and 2D heterostructures[28–30]. Demonstrating a silicon-native SPI system at a longer SWIR wavelength would therefore provide a direct assessment of whether hyperdoped silicon can satisfy the rigorous requirements for practical room-temperature operation using a stable, CMOS-standard platform.

Here, we seek to validate the practical viability of our hyperdoped silicon photodetector for room-temperature SWIR applications by employing computation-based SPI with 1550 nm source as a system-level test platform. We specifically investigate the impact of ultrafast laser heating (ULH), a processing step that has shown significant promise for improving material-level properties such as crystallinity, sub-bandgap absorption and its thermal stability[31–34], but has yet to be validated in a functional device architecture. To determine the optimal fabrication route, we perform a comparative study of devices processed with and without ULH using 1550 nm laser characterization as the deciding benchmark. The optimized device is then subjected to a comprehensive suite of photoelectrical, dynamic, and noise characterizations. Finally, we co-optimize the detector biasing conditions and readout electronics, recognizing that system-level performance in SPI and related applications depends critically on the interplay between intrinsic device properties and external signal amplification and acquisition. Through this structured approach, we assess the potential of a silicon-native, CMOS-compatible platform for room-temperature SWIR photonics.

## Room-temperature SPI with hyperdoped silicon photodetector

**Figure 1a** depicts a recently developed computation-based SPI scheme, here adapted by replacing the detector with our hyperdoped silicon device (See ref[29] and **Supplementary Text 1** for details). In short, the light source illuminates the object, and the transmitted light is re-imaged onto a commercial Digital Micromirror Device (DMD, Vialux DLP7000). The DMD comprises an array of independently addressable micromirrors that can be switched between on and off states, enabling computer-controlled spatial modulation of the reflected light through sequentially applied patterns. For each pattern, the total reflected intensity is recorded by the single-pixel hyperdoped silicon detector, and the image is reconstructed numerically from the known modulation sequence and the corresponding detector signals.

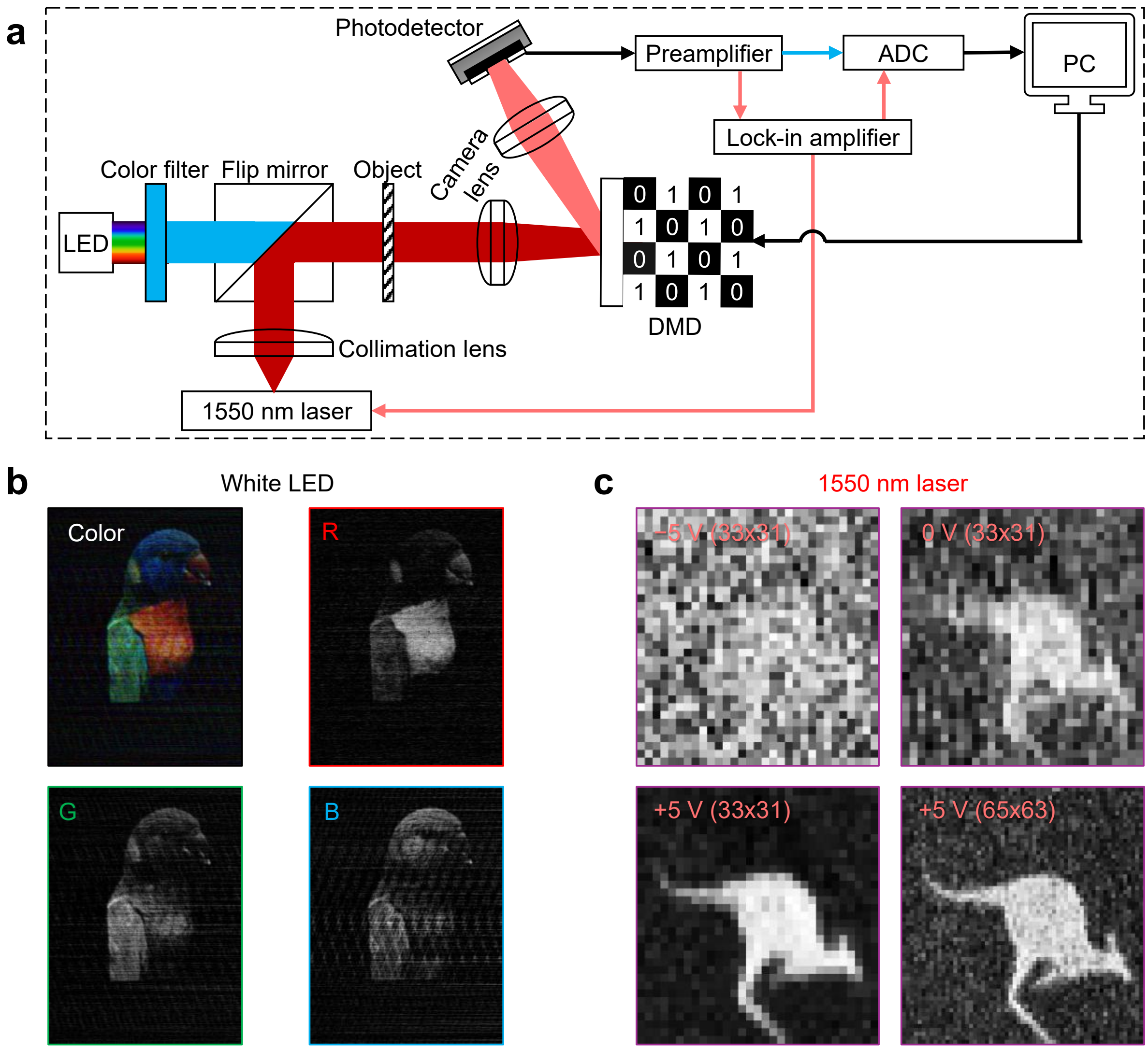

**Figure 1 | Single-pixel imaging (SPI) enabled by a sulfur-hyperdoped silicon photodetector. a,** Schematic of the SPI system. The red path indicates the optical and electronic signal route for 1550 nm laser illumination, while the blue path corresponds to broadband LED illumination. The object is spatially modulated by a digital micromirror device (DMD), and the reflected light is collected by a sulfur-hyperdoped silicon photodetector, followed by pre-amplification, lock-in detection (for 1550 nm modulation), analog-to-digital conversion (ADC), and computer-based reconstruction. **b,** Multispectral SPI of a rainbow lorikeet object under white-LED illumination, showing the reconstructed full-color image and the corresponding red (R), green (G), and blue (B) channels with a resolution of 768 × 1023 pixels. **c,** Reconstructed images from a kangaroo shaped object obtained under 1550 nm laser illumination at different bias voltages and spatial resolutions, demonstrating the dependence of image quality on operating conditions.

To first validate the functionality of our hyperdoped silicon detector with the imaging setup, we perform multispectral imaging with a visible light source to leverage the high intrinsic absorption of silicon for an initial optical calibration. A semitransparent rainbow lorikeet target is illuminated with a white-light LED, and red, green and blue filters are sequentially inserted to enable wavelength-selective measurements. Owing to the inherently large above-bandgap response, the signal-to-noise ratio is high enough to ensure no special optimization, except the usual optical path alignment, is required to realize imaging. In this regime, the photocurrent is converted to a voltage by a preamplifier and directly digitized, while spatial patterns are projected at the DMD’s maximum operating frequency (~22.7 kHz). **Figure 1b** shows reconstructed RGB images with 768 × 1023 pixels from an acquisition time of 47 s for each panel, demonstrating that the large-area hyperdoped-Si detector supports high-resolution, high-speed, multispectral SPI when light absorption is not bandgap-limited.

Having established system-level functionality, we next evaluate performance at a SWIR wavelength of 1550 nm; while of course central to telecommunications, where eye safety and fiber/atmospheric transparency are paramount, it serves here as a critical quantitative proxy for the broader SWIR spectrum. In this case, the object is illuminated with a 1550 nm laser, 1.5 kHz intensity-modulated for lock-in detection, and reconstructed from sequential patterned measurements. As shown in **Figure. 1c**, we observe the imaging quality is bias-dependent: under reverse bias (−5 V) the object is barely discernible, and at zero bias the reconstruction remains blurred. Surprisingly, a forward bias of +5 V yields a markedly sharper image with 33 × 31 pixels.

Increasing the sampling density to 65 × 63 pixels further improves the reconstruction quality, with the kangaroo image becoming better-defined. The optimized image is reconstructed from 4096 patterns with a total acquisition time of 14 min, suggesting that SWIR SPI can be realized within experimentally practical timescales and indicating the hyperdoped detector satisfies certain bandwidth requirement in sub-bandgap response. With the practical viability of the silicon-based SWIR imager demonstrated, we next focus on fundamentally understanding how it works, systematically linking the macro-scale imaging performance back to its micro-scale material fabrication and localized electronic transport.

## Device fabrication and electrical–optical characteristics

**Figure 2a** shows the process flow for the fabrication of hyperdoped silicon detectors. As our lightly doped p-type substrate is hyperdoped with sulfur, nominally generating a heavily doped $n^+$ type region[33], the final device architecture (**Figs. 2b and 2c**) incorporates dual-mode geometry: a vertical configuration (i.e., front-to-back contacts) for primary operation in nominal photodiode mode and a lateral configuration (i.e., front-to-front contacts) for photoconductor mode as well as for verifying the establishment of the ohmic contacts.

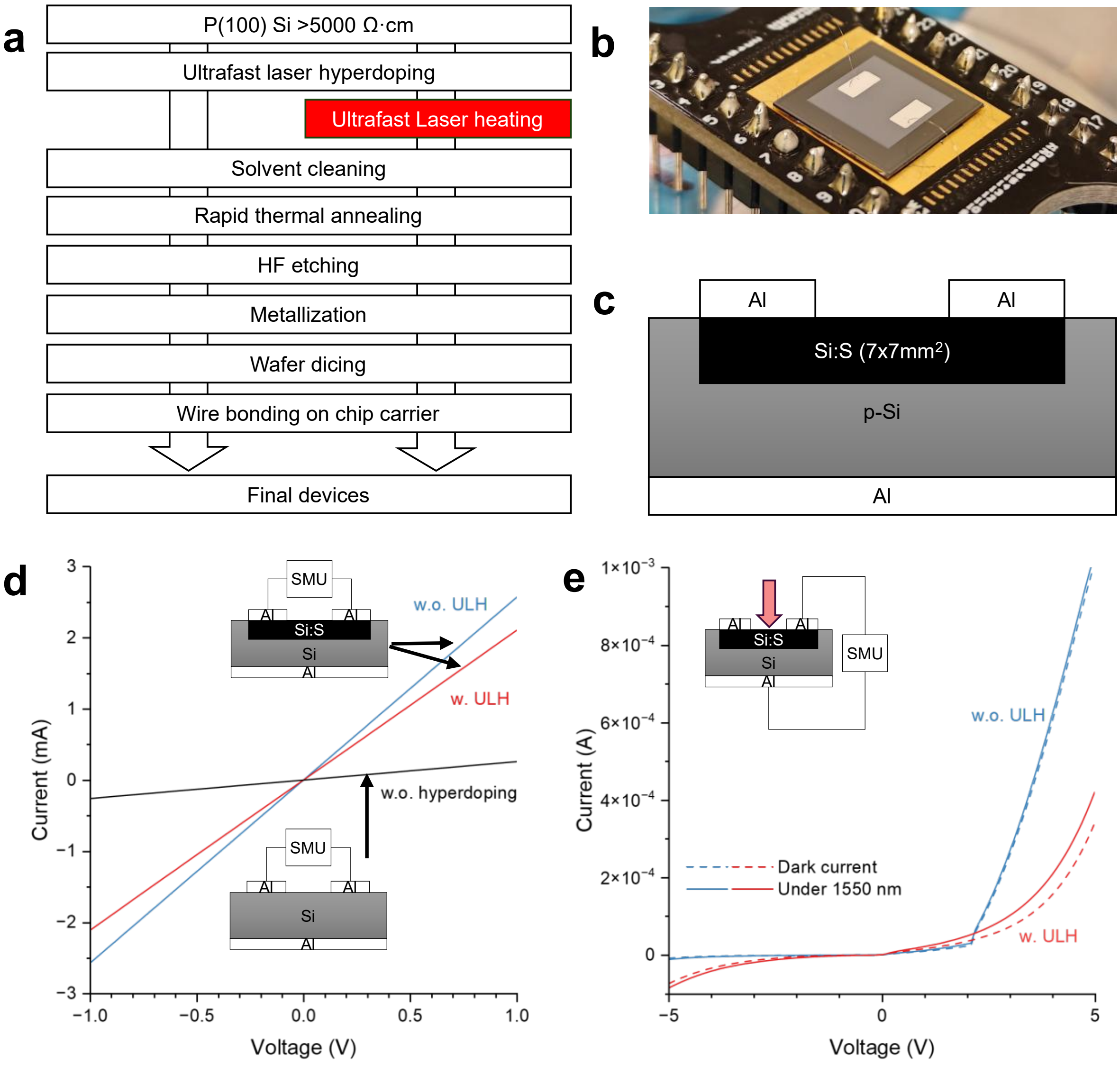

**Figure 2 | Fabrication process, device architecture, and electrical–optical characteristics of the sulfur-hyperdoped silicon photodetectors. a**, Process flow for device fabrication on high-resistivity p-type Si, including ultrafast-laser sulfur hyperdoping with optional ultrafast laser heating (ULH). **b**, Photograph of a representative packaged device mounted on a chip carrier and prepared for electrical characterization. **c**, Schematic cross-section of the photodetector with nominal $n^+$ type Si:S layer. **d**, Representative current–voltage (I–V) characteristics under dark measured between front-front contacts. **e**, Dark and light (1550 nm, 4.88 mW) illuminated I–V curves for devices fabricated with and without ULH. Inset shows a schematic of the measurement between front-rear contacts.

Initially, we fabricated a series of detectors with systematically varied laser fluence and scan speed during ultrafast laser hyperdoping and introduced an additional ULH step (**Fig.2a**) for a subset of samples to assess its impact at the device level. To strictly isolate the influence of the ULH treatment on device performance, our analysis focuses on a representative pair of detectors showing sub-bandgap response in vertical mode (**Supplementary Text 2**). Both devices were processed under completely identical hyperdoping conditions, with the ULH step serving as the single experimental variable.

Following wafer dicing and packaging, we verified the contact behavior via lateral I–V measurements using a source measurement unit (SMU). **Figure 2d** shows that all samples including ULH-treated, a non-ULH, and an additional pristine-Si reference (to verify the back contact) exhibit linear, ohmic characteristics. Notably, the ULH-treated sample displays a reduced I–V slope relative to the non-ULH counterpart, originating from an increase in sheet resistance or a decrease in sheet carrier density following the ULH, in line with previous material study[33,35].

Having established the sub-bandgap activation, we next address the highly unconventional bias dependence observed during the SPI experiment (**Fig. 1c**). Given that the optimal imaging performance occurs under forward bias, a behavior entirely unexpected for a standard p–n junction, a further investigation into the bias-tunable transport mechanisms is essential. Therefore, we characterized the I–V response in the dark and under 1550 nm illumination (**Fig.2e**).

The measured curves exhibit weak rectification consistent with the formation of a laser-induced p-n junction. Notably, the ULH-treated device shows a slight reduction in rectification compared to the non-ULH counterpart, characterized by a marginally lower forward dark current, consistent with horizontal mode (**Fig.2d**), along with a higher reverse dark current. The reduction in (forward) dark current under both modes suggests that the ULH step modifies the effective junction properties, potentially through a redistribution of active dopants or a change in the trap-assisted transport mechanisms that govern sub-bandgap carrier collection. Interestingly, under 1550 nm illumination, the non-ULH device exhibits a marginal change in current, while the ULH-treated device shows a pronounced photocurrent response, particularly under forward bias.

Crucially, the higher photocurrent at +5V bias aligns with our experimental observations from the sub-bandgap SPI demonstration on the same ULH-treated device, where this bias yielded the highest reconstruction fidelity and contrast. Previous studies[31–33,36] indicate that millisecond-scale ULH processing can, on the one hand, preserve or reactivate sub-bandgap absorption by driving sulfur dopants into optically active configurations that remain stable under subsequent thermal treatment. On the other hand, ULH also suppresses dark-current-related noise by improving crystallinity and reducing leakage pathways in the hyperdoped region (See **Supplementary Text 3** for detailed discussion). Accordingly, the sub-bandgap imaging performance and increased photocurrent we observe demonstrate that ULH provides a practical defect-engineering strategy for silicon-native SWIR photodetectors, enabling an optimized trade-off between sub-bandgap absorption and dark current noise under forward bias.

However, these SMU-based measurements have inherent limitations. As the bias voltage increases and the dark current grows, the large DC background forces the SMU into a high-current range where the sub-bandgap photocurrent approaches the measurement uncertainty, particularly when the optical excitation is weak (e.g., reduced laser intensity or lamp–grating sources). This issue can be further escalated by unshielded, random ambient light that effectively adds to the dark current (**Supplementary Text 4**). Consequently, the historically modest performance reported for hyperdoped silicon devices at longer SWIR wavelengths may largely reflect these measurement bottlenecks and the use of non-optimized front-end readout circuits, rather than intrinsic material limits. This motivates a comprehensive dynamic (laser-modulated) characterization and co-optimization of the readout electronics to accurately extract the intrinsic performance of the ULH-treated device and thereby realize the detector's full potential.

## Room-temperature 1550-nm dynamic photoresponse

As demonstrated in our SPI system, the low-noise current preamplifier is a fundamental prerequisite for converting the photocurrent into an ADC-readable voltage signal, allowing us to access the high-quality image with 1550 nm from the detector. As visualized in **Figure 3a**, the preamplifier provides a tunable bias (up to ±5 V) and adjustable transimpedance gain (reciprocal to sensitivity), and together with the laser modulation reference, the output voltage is recorded and translated to photocurrent using the known gain. To prevent the large bias-induced dark current from overloading the readout electronics, a DC offset must be applied so that the weaker transient photocurrent can be accurately amplified and resolved.

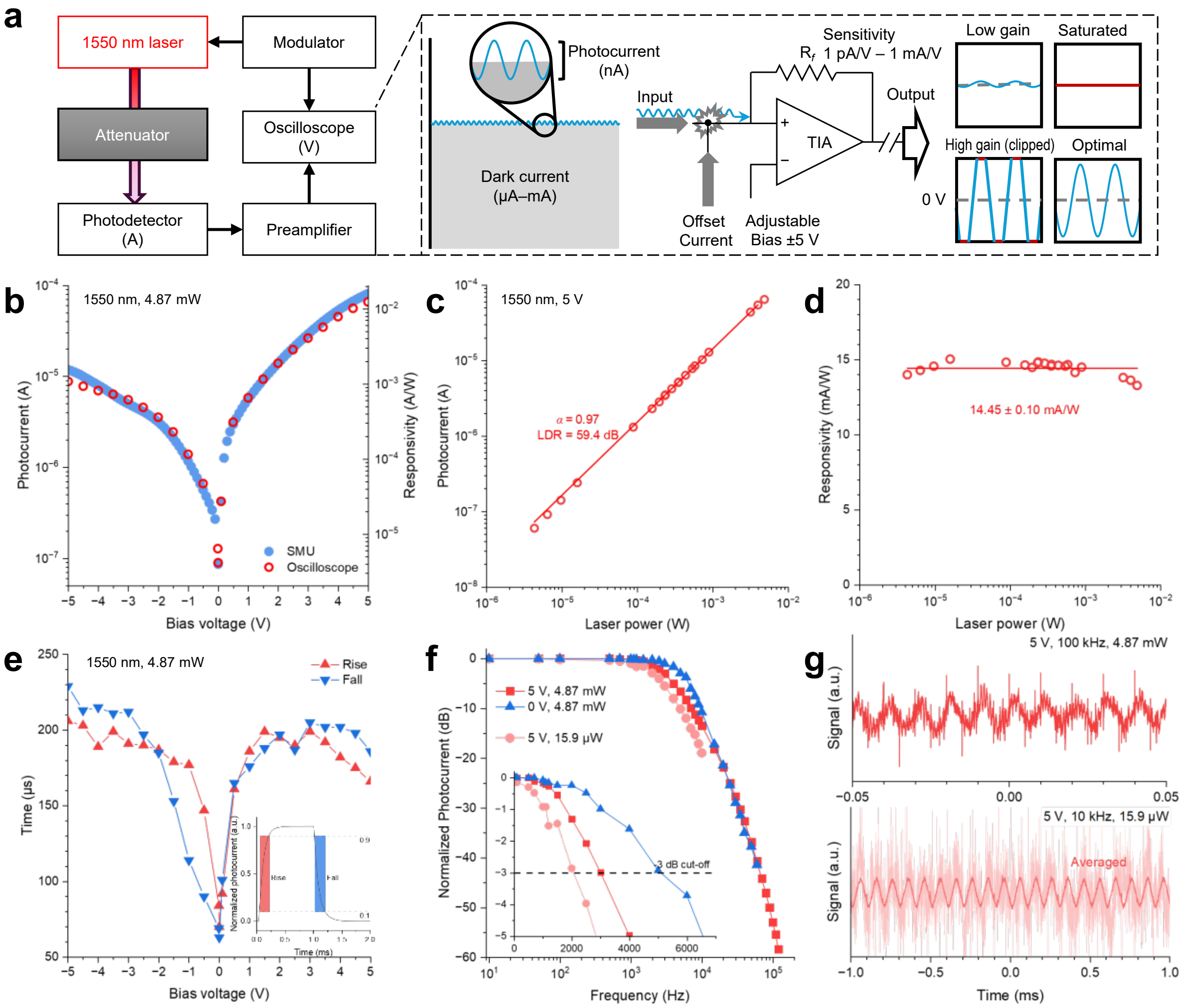

**Figure 3 | Dynamic electrical-optical characterization of the sulfur-hyperdoped silicon photodetector at 1550 nm. a**, Measurement setup for 1550-nm dynamic characterization. Inset schematic shows the photocurrent is amplified by a low-noise transimpedance amplifier (TIA) with adjustable sensitivity and bias, followed by signal and output amplification and

oscilloscope readout. All measurements are performed under vertical mode. **b**, Bias-dependent photocurrent and responsivity measured under 1550-nm illumination (4.87 mW) with a comparison of data collected from IV (under continuous illumination) and TIA + oscilloscope (light modulated at 500 Hz). **c**, Photocurrent as a function of incident laser power at 5 V bias, exhibiting linear behavior over nearly three orders of magnitude and yielding a linear dynamic range (LDR) of 59.4 dB. **d**, Extracted responsivity versus optical power, showing a nearly constant average value of 14.5 mA $W^{-1}$ at 1550 nm. **e**, Rise and fall times as a function of bias voltage measured at 1550 nm (4.87 mW), with representative transient traces (at 5 V) shown in the inset. **f**, Frequency-dependent normalized photocurrent at different bias voltages and optical powers, indicating −3 dB cutoff frequencies in the kHz range. **g**, Time-domain photocurrent signals recorded at 5 V under high-frequency modulation (100 kHz, 4.87 mW) and low-power operation (10 kHz, 15.9 μW), with averaged traces highlighting stable periodic response.

**Figure 3b** shows the photocurrent obtained from our SPI-ready preamplifier setup matches the steady-state results derived from SMU measurements ($I_{1550\ \mathrm{nm}} - I_{\mathrm{dark}}$ in **Fig. 2e**) over ±5 V bias range. The divergence observed at larger biases is likely due to thermal effects; specifically, continuous illumination during SMU sweeps can trigger self-heating, resulting in an artificially inflated conductive change that is absent during transient, modulated operation. Note these comparisons were performed at maximum available laser intensity to ensure the photocurrent remained within the detectable range of the SMU. However, at the lower power densities typical of SWIR sensing and imaging, the SMU would fail in accurate measurement and thus an optimized preamplifier setup could allow high-fidelity signal extraction even when the photocurrent is several orders of magnitude weaker than dark current.

Importantly, the photocurrent is systematically larger under forward bias than under reverse bias, mirroring the behavior of the dark current in **Fig. 2e**. This indicates predominantly photoconductive operation at 1550 nm rather than efficient photovoltaic junction behavior. This forward-biased optimum is consistent with several reports on hyperdoped-Si photodetectors with similar vertical architecture, but contrasts with most other studies where reverse bias yields the best performance (**Supplementary Table 1**). These discrepancies suggest that the optimal operating polarity is highly sensitive to the detailed junction configuration and defect distribution in the hyperdoped region, which cannot be treated as an ideal photodiode or photoconductor.

For sub-bandgap detection, a high dopant density near the surface is essential to enable sufficient absorption and carrier generation[18,19]. However, in our structure the hyperdoped region confined near the surface, while the depletion region of the nominal p–n junction primarily extends into the lightly doped substrate. Consequently, photocarriers generated within the hyperdoped layer lie largely outside the space-charge region and therefore do not directly benefit from the enhanced carrier separation typically provided by reverse bias. For simplicity, the device can thus be viewed as a hyperdoped conductive layer electrically in series or partially overlapped with a relatively weak rectifying junction[33], although a comprehensive theoretical treatment of these sub-bandgap carrier dynamics is beyond the scope of this demonstration and warrants dedicated future study. Under forward bias, the junction field is partially reduced, and carrier transport within the hyperdoped region becomes dominant, leading to behavior more characteristic of a photoconductive mode in vertical direction rather than conventional depletion-region photodiode operation. Under +5 V bias, the detector achieves an apparent linear dynamic range of 59.4 dB (**Fig. 3c** and **Supplementary Text 5**) corroborated by an average responsivity of $R \approx 14.5$ mA W$^{-1}$ at 1550 nm (**Fig. 3d**).

Detector dynamics are critical for SPI where the modulation rate sets the achievable imaging speed. Under 1550 nm modulation, both rise and fall times are longer under bias than at 0 V (**Fig. 3e**). These quantities are extracted from the averaged temporal traces as exampled in **inset of Fig. 3e**. Prior reports on hyperdoped silicon photodetectors with similar architectures highlight the role of defect-related trapping centers in governing photoconductive gain mechanism[18,19,24], which increases the responsivity but lowers the response speed. Here we further clarify that the bias-activated gain should originate primarily from an imbalance between mobile majority and minority carriers within the conduction channel[37]. Upon illumination, photogenerated minority carriers are preferentially localized by defect and surface-related trap states, while their majority counterparts remain mobile and accumulate in the conductive hyperdoped layer. This imbalance enhances channel conductivity and results in an apparent photoconductive gain. Thus, the device transits from a photodiode mode to a photoconductive mode under forward bias. Increasing the bias could modify the apparent minority carrier lifetime presumably by altering carrier confinement, trap occupation, and the local electric-field distribution, thereby tuning both the magnitude of the gain and the detector dynamics.

Forward bias at +5 V offers a favorable compromise between photoconductive gain and speed and is therefore preferred for high-speed imaging. This is in contrast with previous studies showing higher photocurrent gain with the expense of milliseconds response time[24,27,38]. The modulation bandwidth, obtained from the photocurrent as a function of modulation frequency (**Fig. 3f** and **Supplementary Text 6**), yields −3 dB cutoff frequencies $f_c$ of ~5 kHz at 0 V and ~3 kHz at +5 V under 4.87 mW illumination, in reasonable agreement with the values inferred from the temporal response (~5 kHz and ~2.1 kHz, respectively; **Supplementary Text 7**).

For SPI operation, the optical power incident on the detector is substantially reduced by object attenuation, DMD modulation and relay-optic losses[29]. Analogously, at a substantially reduced incident power of 15.9 μW and +5 V bias, the $f_c$ decreases to ~2 kHz (**Fig. 3f**), but remains on the same order of magnitude, indicating only weak dependence of the temporal response on illumination in the relevant range.

The −3 dB frequency does not represent a hard operating limit: modulated signals remain resolvable well above $f_c$ (**Fig. 3g**). Under strong illumination, a clear periodic response is observed up to 100 kHz, where the preamplifier bandwidth becomes limiting, and at low power the signal remains visible at 10 kHz with temporal averaging. Thus, under realistic SPI conditions, the device supports modulation rates comfortably within the kHz regime with headroom for operation at higher frequencies via device and optical engineering.

## Readout co-optimization and noise-limited detectivity

Since practical operation is determined by signal-to-noise rather than responsivity alone, to enable a fair comparison of device performance under realistic operating conditions, we extract the specific detectivity $D^*$ as a practical figure of merit[39,40]. To ensure a scientifically valid measurement of $D^*$, it is critical that the photocurrent and the noise spectral density are characterized under the same conditions.

**Figure 4a** plots the output voltage as a function of preamplifier sensitivity $S$ for high (4.87 mW) and low (15.9 µW) incident powers. In the high sensitivity regime, reducing $S$ leads to a linear increase in output voltage while the inferred photocurrent, i.e., amplitude × sensitivity, remains essentially unchanged. At smaller $S$, the response becomes power-dependent. At $S$ < 20 µA V⁻¹, the output under 4.87 mW illumination overloads, indicating that the preamplifier input or output stage has reached its dynamic range limit. Under 15.9 µW illumination, the response begins to deviate from linear trend in this regime ($S$ < 20 µA V⁻¹). This is because the effective bandwidth decreases at higher gain, so the actual gain at the 500 Hz modulation frequency starts to be lower than the nominal value.

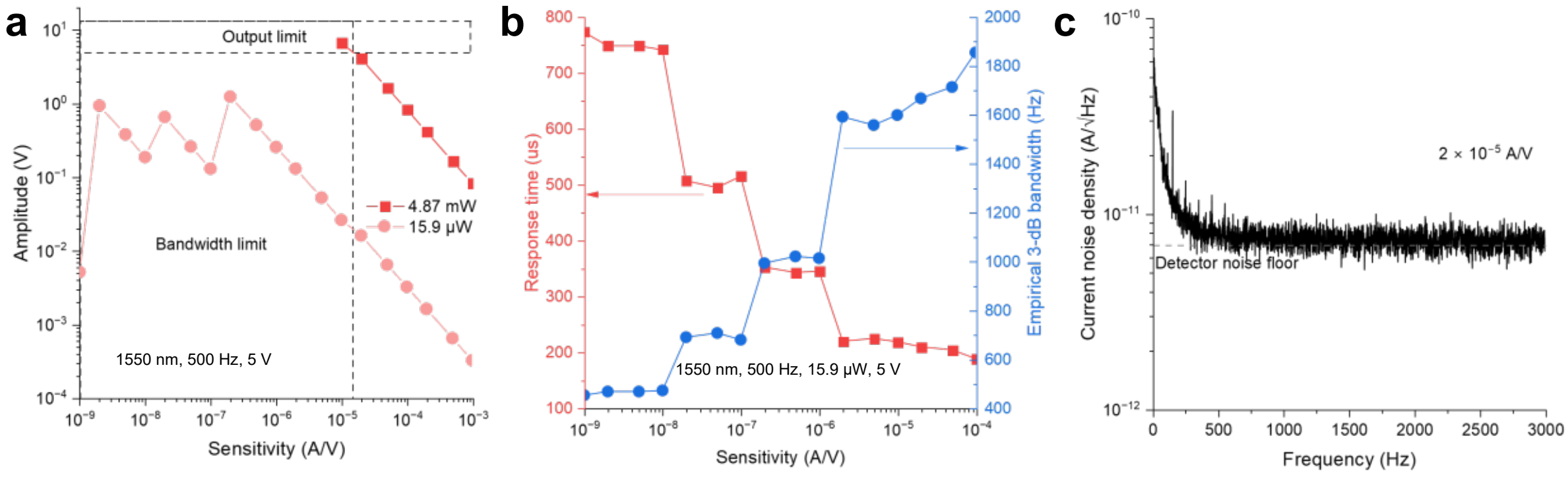

**Figure 4 | Preamplifier optimization and noise analysis of the sulfur-hyperdoped silicon photodetector. a,** Measured voltage amplitude at the amplifier output as a function of transimpedance-amplifier sensitivity for two incident optical powers (4.87 mW and 15.9 µW at 1550 nm) at a bias of 5 V. **b,** Extracted temporal response time (left axis, averaged rise and fall times) and nominal bandwidth (right axis) as a function of amplifier sensitivity at 1550 nm and 5 V bias. **c,** Spectral current noise density measured at sensitivity of $2 \times 10^{-5}$ A V⁻¹ with the estimated shot and thermal noise limit at 5 V calculated from dark I–V measurement.

In addition, the impact of $S$ on detection speed is quantified by extracting the response time as a function of $S$ under low-power illumination, since high-power operation saturates at large gains and does not provide reliable timing information. **Figure 4b** shows that as $S$ decreases, the response time increases with corresponding step-like changes from discrete internal gain ranges of the preamplifier, and the effective cutoff frequency $f_c$ drops from the kHz range at low gain to the sub-kHz range at the highest gains. This suggests that an intermediate gain is required to amplify the photocurrent with little input-referred noise, to avoid overload and to preserve the bandwidth, thereby realizing the practical functionality of the hyperdoped silicon detector in SWIR applications.

We next characterize the noise of the detector–preamplifier system using FFT-based current-noise spectra (see **Supplementary Text 8** for details). **Figures 4c** shows the current-noise spectral density $i_n(f)$ at the sensitivity $S$ = 20 µA V⁻¹ and bias of +5 V, together with the shot and thermal noise floor estimated from dark I–V curve (**Supplementary Text 9**). At low frequencies, the spectrum exhibits a pronounced 1/$f$-like slope and possibly mixed with trap-related generation–recombination noise, while in the 500–3000 Hz range it flattens into a white-noise plateau that approaches the detector noise floor, dominated by the shot noise from relatively large dark current. For practical operation, the laser modulation frequency should therefore be chosen above the 1/*f*-dominated region, where the noise is close to the white-noise limit that is dominated by the shot noise.

The specific detectivity $D^*$ is extracted from the measured photocurrent (**Fig. 4a**) and the average white noise $\bar{\imath}_n$ over 500–3000 Hz (**Fig. 4c**). Under a +5 V bias and 1550 nm illumination, we obtain $D^* \approx (1.91 \pm 0.18) \times 10^9$ Jones (or $\mathrm{cm\,Hz^{1/2}\,W^{-1}}$) for 15.9 μW and $D^* \approx (1.59 \pm 0.16) \times 10^9$ Jones for 4.87 mW. These values lie near the upper bound of reported performance for silicon-native photodetectors at 1550 nm without heterojunction integration (**Supplementary Table 1 and Ref**[26]). By contrast, $D^*$ estimated at 0 V and −5 V is $3.26 \times 10^8$ and $3.33 \times 10^8$ Jones, respectively, using the photocurrent in **Fig.3b** and dark I–V inferred noise floor. The highest detectivity at +5 V bias is consistent with the best image quality achieved at this forward-bias operating point.

## Conclusion

These results represent the first demonstration of room-temperature SWIR computational imaging using silicon-native architecture. The large-area sulfur-hyperdoped silicon detector, engineered by ultrafast laser processing with a ULH step, delivers responsivity of ~14.5 mA $W^{-1}$, specific detectivities above $10^9$ Jones, a linear dynamic range of ~59 dB and kHz-scale bandwidth, with ULH improving the photocurrent while suppressing the dark current resulting in a forward bias, photoconductive-like operation mode. As a proxy for the broader SWIR applications, we show the hyperdoped-silicon integrated SPI architecture enables object reconstruction at 1550 nm in addition to high-resolution multispectral (RGB) imaging, demonstrating that a single silicon-based pixel can tackle broadband visible to SWIR imaging. The detector-limited noise behavior we observe under matched readout suggests that hyperdoped silicon's SWIR performance has been systematically understated by conventional characterization. Further advances in defect engineering, device design and readout co-optimization could extend silicon's usefulness deeper into the SWIR for telecom, sensing and imaging applications.

## Materials and methods

### *Device fabrication and screening*

The hyperdoped silicon photodetector was fabricated on a 280-μm thick, double-side-polished, p-type, 4″ float-zone (100) silicon wafer with a resistivity of 5000 ± 3000 Ω·cm. The wafer was first irradiated with ultrashort laser pulses in 675 mbar sulfur hexafluoride atmosphere. The Yb:YAG laser source (Amplitude Tangor 100) emitted pulses at the central wavelength 1030 nm with 10 ps pulse duration at an adjustable repetition rate of 200 kHz to 41 MHz. The output pulse energy of the laser source was controlled with an acousto-optical modulator. The laser pulses passed through a scanning system consisting of a two-axis galvanometer scanner and a *f*-theta lens with a focal length of 340 mm and were focused onto the wafer surface inside the processing chamber. The $1/e^2$ beam diameter was measured with a camera-based beam profiling system in the focal plane to be $2w_0 = 70\ \mu$m. The pulses were scanned across the wafer surface in a square-shaped grid pattern with a size of 7 mm × 7 mm with a period of 1 cm × 1 cm. Hyperdoping was performed using 500 pulses per area with a peak fluence of 1.2 J $cm^{-2}$. Subsequently, ULH was applied to selected samples in ambient air using a repetition rate of 41 MHz, a scanning speed of 3000 mm $s^{-1}$, and a peak fluence of 10 mJ $cm^{-2}$. These processing parameters were optimized for the devices used in the SPI demonstration; additional variations are documented in **Supplementary Text 2**.

After laser processing, the wafer was subsequently cleaned in ultrasonically agitated acetone and isopropyl alcohol and then rinsed in de-ionized water to remove the dust residues. We then performed rapid thermal annealing (RTA) at 900 °C for 10 s in an $N_2$ ambient. The wafer was then etched in 1% hydrofluoric acid (HF) for 1–2 min to remove native oxides. Immediately after HF etching, metallization was carried out via e-beam evaporation (Angstrom Engineering, model 09340) of 500 nm Al on both sides of the wafer. The front side was patterned using a 304 stainless steel shadow mask, while the back side received a full-area Al layer. The resulting front-side contacts measured 1.5 mm × 3 mm and were spaced by 3 mm.

To verify Ohmic contact behavior, the devices were measured between two adjacent front contacts using a semiconductor parameter analyzer (Keysight/Agilent 4156A). To confirm Ohmic behavior on the back planar side, a reference sample with two pads on the non-hyperdoped, pristine area was also fabricated and characterized.

After initial testing of sub-bandgap responsivity, the wafer was diced into 1 cm × 1 cm devices using a UV laser (LPKF U4) and selected devices were then individually bonded onto custom chip carriers for device characterization and imaging experiments. To compare devices with and without the ULH step, each chip was mounted in a light-shielded enclosure, and I–V characteristics were measured by a SMU (Keysight 2450) under dark conditions and under continuous 1550 nm laser illumination, with a maximum intensity of 4.87 mW calibrated by a Ge photodiode (Thorlabs FDG03).

***Photocurrent characterization under laser illumination***

For 1550-nm laser characterization, the device was mounted in the same light-shielded enclosure and illuminated with a 1550 nm laser diode (LD) whose intensity was calibrated by the same Ge photodiode. The LD was driven by a current controller (Thorlabs LDC205C) and modulated by a function generator (Stanford Research DS345) using a square-wave signal at selected frequencies. The device output was pre-amplified with a low-noise current amplifier (Stanford Research SR570) and converted to a voltage signal, which was then recorded with an oscilloscope (Tektronix MSO2024B). A set of neutral density (ND) filters was used to attenuate the laser intensity, enabling measurement of laser-power-dependent photocurrent for linear dynamic range analysis and low-light characterization.

By adjusting the preamplifier sensitivity $S$, we effectively tuned the transimpedance gain and determined the input current $I_{\text{in}}$ from the measured oscilloscope voltage using the known current-to-voltage gain:

$$I_{\text{in}} = V_{\text{out}} \cdot S = \frac{V_{\text{out}}}{G},$$

where $V_{\text{out}}$ is the recorded preamplifier output voltage and $G$ is the transimpedance gain.

**Acknowledgement**

X.L. acknowledges financial support from the Research Council of Finland through the Academy Research Fellowship (#354199). S.S. acknowledges funding by the German Federal Ministry of Education and Research in the context of the federal-state program “FH-Personal” under the grant number 03FHP147A (REQUAS). The work is part of the Research Council of Finland Flagship Program, Photonics Research and Innovation (PREIN, #346529). The authors acknowledge the provision of facilities and technical support by Aalto University at OtaNano-Micronova Nanofabrication Centre. We thank Marko Yli-Koski and Rob Delaney for their invaluable technical assistance.

**Author Contributions**

Conceptualization: XL, SS, SK, KC, JB, HS; Data curation: XL, JC, PMK, SP; Formal Analysis: XL; Funding acquisition: XL, SS, VV, SK, KC, JB, HS; Investigation: XL, SS, JC, PMK, SP, VAV; Methodology: XL, SS, JC, PMK, SP, KC; Project administration: XL, SS, KC, JB; Resources: XL, SS, VAV, VV, SK, KC, JB, HS; Software: XL, JC, PMK, SP, VAV; Supervision: SK, KC, JB, HS; Validation: XL, SS, VV, SK, KC, JB, HS; Visualization: XL, JC; Writing – original draft: XL; Writing – review & editing: XL, SS, JC, PMK, SP, VAV, VV, SK, KC, JB, HS

**Conflict of interest**

The authors declare no competing interests.

**Data availability statements**

The authors declare that all datasets on which the conclusions of the paper reply are openly available within the paper, its supplementary information files, and Zenodo (DOI: 10.5281/zenodo.20326707).

**Supplementary Information for**

# Hyperdoped silicon photodetectors enable room-temperature computational SWIR imaging at 1550 nm

## CONTENTS

# SUPPLEMENTARY TABLE

**Table S1**. Comparison of room-temperature hyperdoped-silicon-based photodetectors operating at 1550 nm (When values at 1550 nm are unavailable, the closest reported wavelength is given in parentheses. J – dark current density, P – light power density, R – responsivity, LDR – linear dynamic range, D* – specific detectivity, $\tau_r/\tau_f$ – rise/fall times).

| Material | Size | Bias | $I_{dark}$ | J (mA/cm2) | P | R | LDR | D* | $\tau_r/\tau_f$ |
|---|---|---|---|---|---|---|---|---|---|
| **Si:S (This work)** | **7×7 mm$^2$** | **5 V** | **146.5 µA** | **0.299** | **4.87 mW 15.9 µW** | **14.5 mA/W** | **59.4 dB** | **$1.59\times10^9$ Jones $1.91\times10^9$ Jones** | **166 µs / 186 µs** |
| Si:Te [1] | 140×450 nm$^2$ (lateral waveguide incident) | −9 V | ≈ 2 µA | - | 125 nW | 0.56 A/W | 33.7 dB | (not able to determine due device structure) | (59 ps from RC estimation) |
| Si:Te [2] | ≈5 mm$^2$ | 0 V | - | - | 20 mW cm$^{-2}$ | 0.3 mA/W | - | $9.2\times10^8$ Jones | 39 µs / 42 µs |
| Si:Ar [3] | 5×5 mm$^2$ | 12 V | ≈ 5 mA | 20 | 3.16 mW | 1.28 A/W | - | ($1.14\times10^{10}$ Jones @1310 nm) | - |
| Si:Ti [4] | 1×1 cm$^2$ | 1 mA | - | 1 | - | 34 mV/W | - | $1.7\times10^4$ Jones | - |
| Si:Se [5] | ≈5 mm$^2$ | −1 V | 57.3 µA | 0.229 | 35 mW | 72 µA/W | ≈ 27.3 dB | - | 7 ns / 23 ns |
| Si:Ti/DLC [6] | 1 mm$^2$ | 0 V | - | - | 63 mW | 2.34 µA/W | - | $1.16\times10^8$ Jones | 5 ms / 3 ms |
| Si:Zn/Gr [7] | - | −3 V – 0 V | < 1 µA | - | 2 mW | 1.6 mA/W | - | $1.56\times10^7$ Jones | 1.4 ms |
| Si:He&Ar [8] | - | 12 V | 12.5 mA | - | - | 350 mA/W | - | $2.78\times10^9$ Jones | (74 µs / 82 µs @1310 nm) |
| Si:S [9] | 5×3 mm$^2$ | −20 V | (7.8 µA @−5 V) | (0.052 @−5 V) | - | 0.8 A/W | - | - | (0.65 ms / 2.13 ms @632 nm) |
| Si:Au [10] | 1×1 mm$^2$ | −5 V | ≈ 1.5 µA | 0.15 | - | 11.6 µA/W | - | - | - |
| Si:Te [11] | - | −2 V | - | - | - | 56.8 mA/W | - | - | - |
| Si:Ag [12] | 7×7 mm$^2$ | −3 V | ≈ > 50 µA | ≈ > 0.102 | - | 65 mA/W | (41.4 dB @1310 nm) | - | (75 µs @1310 nm) |
| Si:Er&O [13] | 200/600 µm donut | 2 V | 5 µA | 0.442 | - | (165 µA/W @1310 nm) | - | ($1.39\times10^9$ Jones @1310 nm) | (14 ms / 12 ms @1310 nm) |
| Si:Ag/Gr [14] | 100 µm channel width (FET) | 0.3 V | ~ 5 mA | - | - | ≈ 92 mA/W (from EQE=7.37%) | (≈ 48.5 dB @1342 nm) | - | (122 µs /131 µs @1342 nm) |

# SUPPLEMENTARY TEXT

## Supplementary Text 1. Single-pixel imaging experiment

While the comprehensive architecture of the same single-pixel imaging (SPI) system is described elsewhere[15], here we outline the specific configurations crucial to the present measurements. The light sources used for imaging included a 1550 nm infrared laser (ThorLabs FPL1009S) and a visible-wavelength LED (Thorlabs MNWHL4) combined with red, green, and blue (RGB) optical filters (whose transmittance spectra were shown in **Fig.S1**). Test objects consisted of a glass slide with a gold film patterned with a kangaroo image and a color photographic slide containing a rainbow lorikeet (a parrot) image.

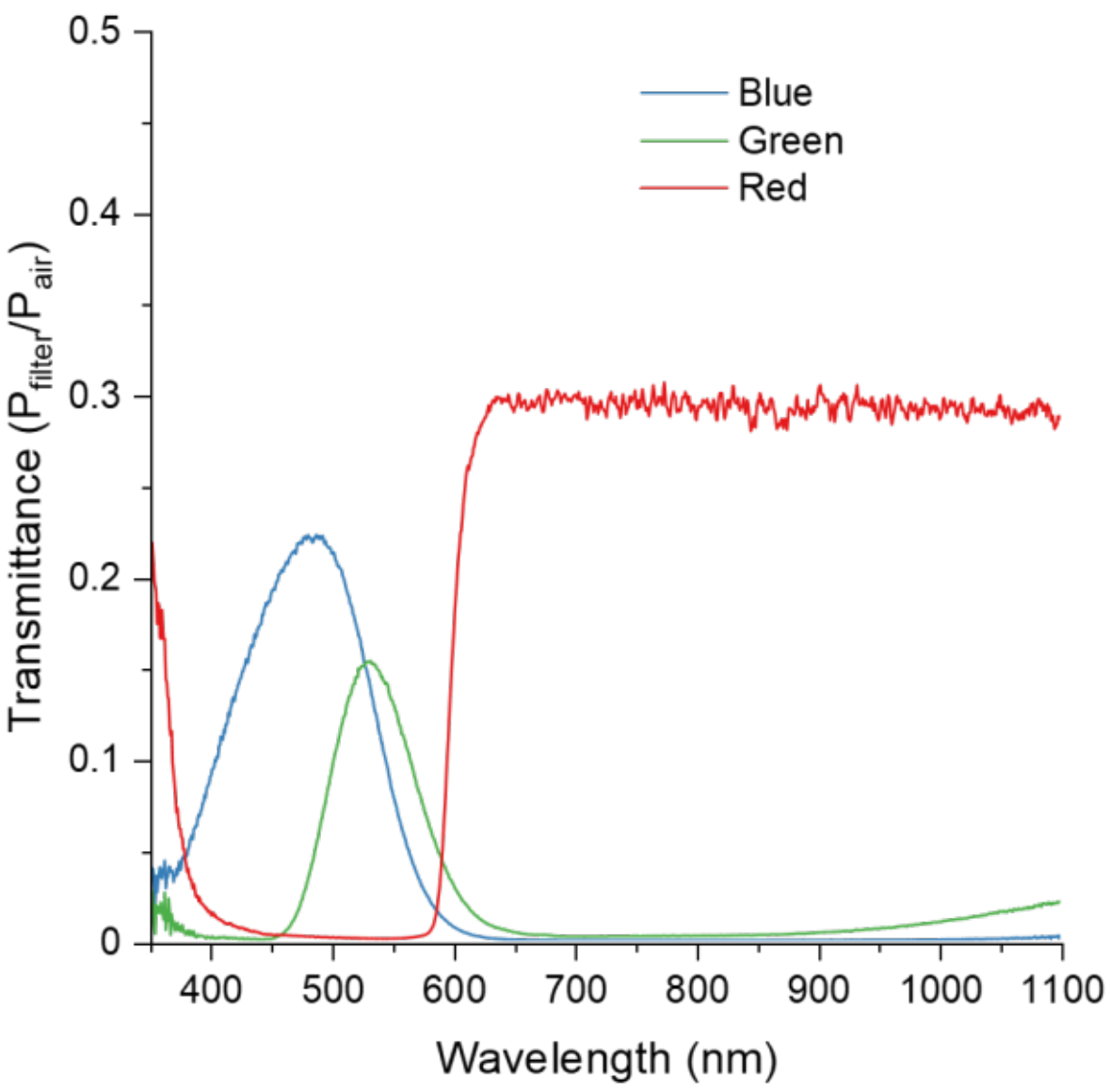

**Fig. S1 |** Measured transmittance spectra of the red, green, and blue optical filters, referenced to air.

As shown in **Fig.1a**, for SWIR imaging, the 1550 nm laser was collimated and directed onto the object via a flip mirror. The transmitted light was imaged by a camera lens (Nikon, focal length 60 mm) onto a DMD (V-7000, Vialux) at normal incidence, which imposed a sequence of spatial modulation patterns. The reflected, pattern-encoded beam was then relayed by a camera lens (Pentax, focal length 50 mm) and a microscope objective (Mitutoyo, 50× magnification) onto the hyperdoped silicon photodetector, which converted the incident intensity into a time-varying electrical signal. This signal was amplified by the preamplifier (which also provided the detector bias) and demodulated by a lock-in amplifier (Stanford Research SR830), with the laser intensity simultaneously modulated at 1.5 kHz. The lock-in amplified output voltage was digitized by a 14-bit analog-to-digital converter (ADC) and transferred to a personal computer for image reconstruction. The sampled voltages $\boldsymbol{V} \in \mathbb{R}^N$ corresponded to the circular convolution of the first row of the cyclic sensing matrix $\boldsymbol{\Phi_1} \in \mathbb{R}^N$, with the imaging object vector $\boldsymbol{O} \in \mathbb{R}^N$, scaled by a normalization constant $m$[16] :

$$\boldsymbol{V} = m * (\boldsymbol{\Phi_1} \circledast \boldsymbol{O}).$$

In our SPI implementation, the cyclic S-matrix patterns were generated using a maximal-length shift-register approach[16], generating a first row of the sensing matrix $\boldsymbol{\Phi_1}$ with a total number pixels of reconstructed image pixels $N = 2^n - 1 = p \times q$, where $n$ was an integer number and $p, q$ defined the dimensions of sampling and image matrices, respectively. During the sampling process, the sensing matrix $\boldsymbol{\Phi_1}$ was first reshaped to a matrix of size $p \times q$ and displayed on the DMD as the first pattern.

Because the sensing matrix was binary, each DMD micromirror turned ‘ON’ when the corresponding matrix element was 1 and ‘OFF’ when it is 0. The ADC then sampled the first voltage value, $V_1$. For subsequent measurements, the DMD displayed the remaining patterns cyclically.

For visible-light (high-resolution) imaging, the preamplifier output was sent directly to the ADC without lock-in detection or light-source modulation. Although the sampling matrix was defined as 1023 × 1025 pixels, the digital micromirror device (DMD) provides only 1024 × 768 micromirrors. Consequently, the full sampling matrix cannot be physically projected, and the reconstructed image resolution was therefore limited to a maximum of 1023 × 768 pixels. A trigger signal was generated by the DMD upon loading each pattern and was sent to the ADC. Upon receipt of this trigger, sampling of the detector signal was initiated by the ADC. The sampled signals corresponding to each displayed pattern were recorded by the control computer.

The reconstruction exploited the equivalence between circular convolution in the spatial domain and multiplication in the Fourier domain. The Fourier transforms of the sampled voltage sequence and of the first row of the cyclic sensing matrix were first computed. The ratio of these spectra was then evaluated in the Fourier domain, followed by an inverse Fourier transform to obtain the reconstructed image.

## Supplementary Text 2. Fabrication and selection of photodetectors

**Table S3** summarizes the hyperdoping parameters and ULH fluence used for all fabricated detectors. Following metallization, we screened all 14 hyperdoped silicon devices across both vertical (front-rear) and lateral (front-front) configurations to identify those exhibiting sub-bandgap photoresponse.

**Table S3**. Laser processing variations and initial responsivity results in front-rear mode for all fabricated devices (*Samples selected for characterization in main text.).

| Sample number | Pulse per spot | Hyperdoping fluence (J cm$^{-2}$) | ULH fluence (mJ cm$^{-2}$) | Responsivity (mA/W) @ 1550 nm, −20 V |
|---|---|---|---|---|
| 01 | 20 | 0.8 | - | <1 |
| 02 | 20 | 0.8 | 13 | <1 |
| 03 | 20 | 0.8 | 14 | <1 |
| 04 | 20 | 0.8 | 15 | 2.0 |
| 05 | 20 | 1.2 | - | 6.4 |
| 06 | 100 | 0.4 | - | <1 |
| 07 | 100 | 0.8 | - | <1 |
| 08 | 100 | 1.2 | - | <1 |
| 09 | 500 | 0.4 | - | 6.6 |
| 10 | 500 | 0.8 | - | 3.9 |
| 11* | 500 | 1.2 | - | 3.3 |
| 12* | 500 | 1.2 | 10 | 3.2 |
| 13 | 500 | 1.2 | 15 | 2.8 |
| 14 | 500 | 1.2 | 20 | <1 |

Photoresponsivity measurements were performed at the wafer level using a commercial system (Bentham PVE300) with the light intensity calibrated by a Ge reference detector (Bentham DH-Ge) over the 800–1800 nm range. The samples were contacted using a probe needle on the front side and an Aluminum foil on the back side. An additional SMU (Keysight 2401) was used to directly contact both ends of the front-side contacts to provide the bias voltage. Following reports that increased reverse bias can enhance photoconductive gain, we performed initial measurements at a reverse bias of 20 V in the vertical configuration. This biasing assumes an n-type hyperdoped region on a p-type substrate. The AC photocurrent signal was pre-amplified by a transformer (Bentham 474) and detected using a lock-in amplifier (Bentham 496).

As shown in **Fig.S2**, while several devices demonstrate clear responsivity beyond the silicon bandgap (> 1100 nm) in the vertical mode (**Fig.S2a**), the signals in the lateral mode remain below the system detection limit (**Fig.S2b**). Furthermore, many devices in vertical mode exhibit a responsivity approximately three orders of magnitude higher at above-bandgap wavelengths (800–1100 nm) compared to 1550 nm. This significant difference explains why the imaging performance in the visible and near-infrared (NIR) regimes remains inherently superior to that achieved at sub-bandgap SWIR wavelengths.

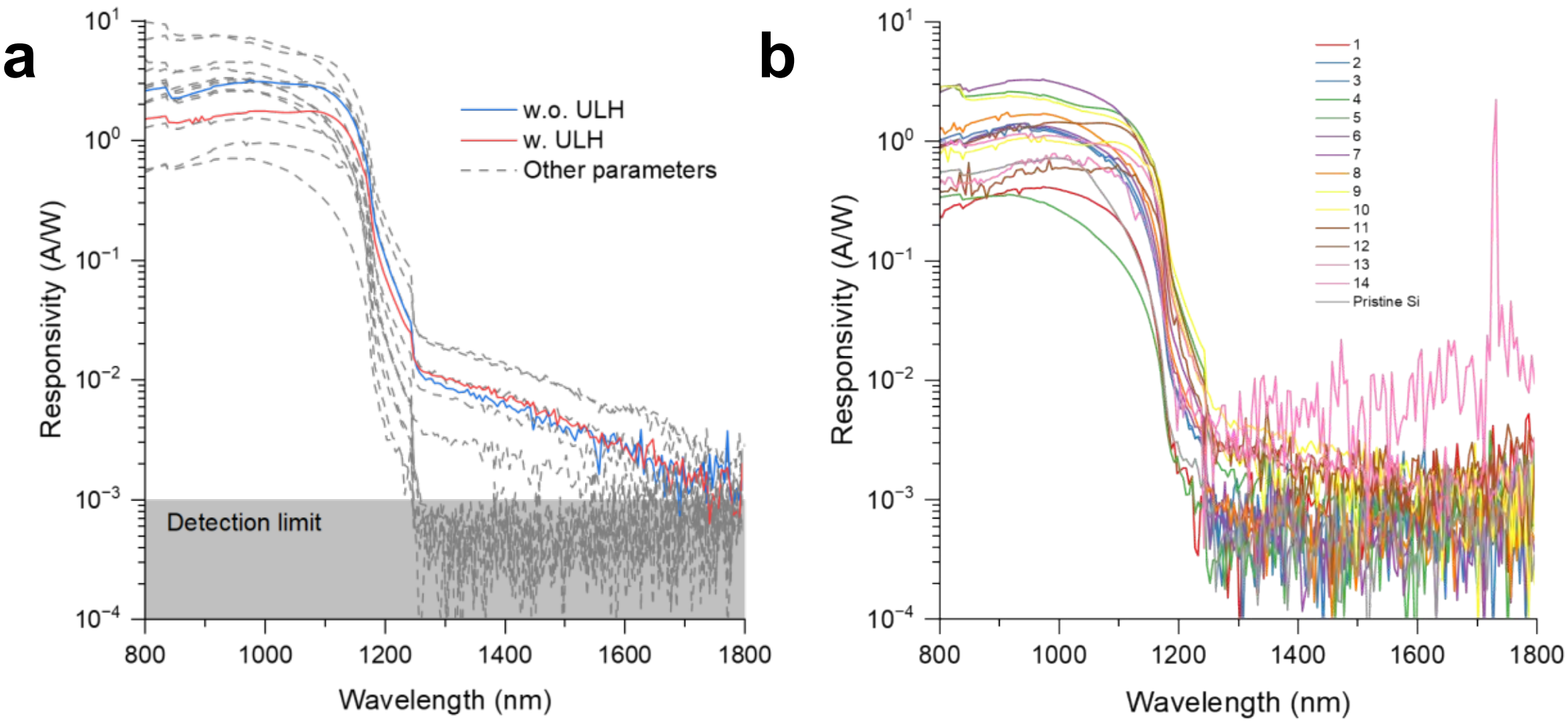

**Fig. S2 | a,** Spectral responsivity measured in vertical mode under −20 V bias for highlighted devices fabricated with (w.) and without (w.o.) ULH, together with additional process-parameter variations. **b**, Spectral responsivity measured in lateral mode showing no sub-bandgap responsivity with a pristine Si device used as a reference. Data was acquired under a 20 V bias, except for device 4, which was evaluated at 5 V to prevent saturation from excessive dark current.

In lateral mode (**Fig.S2b**), the erratic fluctuations observed in the sub-bandgap regime, particularly the anomalously high values for device 14, are not intrinsic material responses. Rather, they are instrumental measurement artifacts occurring where the noise floor of the setup overwhelms the weak intrinsic photocurrent. The lateral configuration fails to exhibit a sub-bandgap photoresponse mainly due to an overwhelming parasitic dark current flowing through the highly conductive surface layer.

By transitioning to a vertical architecture, the extraction path is redirected across the junction. This not only bypasses the conductive surface shunt but also drastically shortens the carrier transit distance, enabling efficient collection across a large active area. To isolate the influence of the ULH step on device performance, we focused our analysis on a representative pair of detectors fabricated under identical hyperdoping conditions, differing only by the application of the ULH treatment (highlighted in **Fig.S2a**). Although these specific devices do not represent the peak responsivity achieved in this study, they provide a controlled platform to decouple the effects of ULH from other processing variables.

## Supplementary Text 3. Origin of improved performance in ULH-processed device

The improved device-level performance (higher sub-bandgap photocurrent and lower dark current, particularly at +5 V) by the ultrafast laser heating (ULH)-processed detector in this work can be understood in the context of the previous systematic study of ULH[17–19] on Si:S hyperdoped layers.

Ultrafast laser hyperdoping of Si in chalcogen-containing atmospheres produces a highly non-equilibrium near-surface region with a substantial fraction of amorphous and heavily disordered silicon and very high local sulfur concentrations in the range of 1 at.%[20]. Raman measurements in prior work show that the as-hyperdoped state contains pronounced amorphous-Si signatures in the 50–200 $cm^{-1}$ and 470–490 $cm^{-1}$ ranges, originating from the rapid resolidification of the laser-induced melt[18]. These amorphous phases are correlated with strong sub-bandgap absorption but also with increased defect scattering and dark (leakage) currents[17]. An "ideal" post-hyperdoping treatment should therefore reduce the non-crystalline fraction while preserving the high sub-bandgap absorptance.

At the atomistic level, defect-engineering models for sulfur-hyperdoped silicon attribute sub-bandgap absorption mainly to deep-level sulfur monomers, which introduce energy states deep within the bandgap and enable sub-bandgap transitions, but do not supply significant free carriers (a source of dark current) at room temperature[17,21–23]. In contrast, shallower sulfur dimers and clusters introduce states closer to the band edges, so they contribute free carriers and dark conductivity but little or no sub-bandgap absorption[17]. Different fabrication and annealing routes are therefore expected to redistribute sulfur among monomer, dimer and cluster configurations, and in doing so jointly tune both the sub-bandgap absorptance and the dark conductivity. Within this framework, an optimal post-processing scheme is one that maximizes the fraction of optically active monomers while minimizing dimer/cluster-related leakage paths.

Systematic post-treatment studies have shown that this balance is delicate. Many purely thermal or top-down etching routes do increase crystallinity, but often at the cost of a noticeable reduction in sub-bandgap absorptance[20], consistent with sulfur diffusion and clustering. By contrast, the combination of controlled thermal diffusion and ULH has been found to move the material into a particularly favorable regime. Previous work[17] shows after a thermal-diffusion step, all ULH-processed samples exhibit higher sub-bandgap absorptance than the as-lasered state, and even thermally "deactivated" samples can have their sub-bandgap absorption reactivated by a subsequent ULH step. At the same time, the same study identifies a group of ULH-treated samples ("Group 2") with comparatively low and thermally stable sheet carrier densities and high carrier mobilities that change only weakly with further diffusion. Mobility analysis indicates that both a low amorphous-silicon fraction and a moderate charge-carrier density are crucial to achieving these high mobilities: reducing amorphous phases lowers defect scattering, while avoiding excessively high doping mitigates free-carrier scattering and dark conduction. On the defect level, these trends are consistent with ULH converting a significant fraction of dimers and clusters into monomers, thereby increasing the monomer concentration and stabilizing the optical and electrical properties against subsequent thermal processing.

The ULH-processed device in this work is designed to operate in the same parameter regime as the favorable ULH-treated group reported previously: it retains strong sub-bandgap absorption around 1550 nm, but with a more crystalline hyperdoped layer, a reduced amorphous fraction and a moderate carrier density, which together yield higher mobility and lower leakage[17] but lower rectification ratio (Fig.2f) with less effective donor density. In contrast, the non-ULH device is expected to preserve more of the as-hyperdoped properties, i.e. a larger amorphous fraction and/or higher free-carrier density, and a less favorable distribution of sulfur among monomer, dimer and cluster states, leading to higher dark current in forward bias, stronger defect-mediated noise, and reduced effective mobility.

Therefore, these material-level trends may provide a consistent explanation for the device-level results reported here: the ULH step drives the hyperdoped region towards a configuration with (i) strong sub-bandgap absorption, (ii) improved crystallinity and reduced amorphous content, and (iii) moderate carrier density with high mobility. In terms of defect chemistry, this corresponds to a higher fraction of deep-level sulfur monomers and fewer dimer-/cluster-related leakage paths. This combination lowers dark-current-related noise without sacrificing responsivity, so that under forward bias (photoconductive mode) the ULH-processed detector achieves a more favorable balance between gain and noise, and thus higher specific detectivity.

## Supplementary Text 4. Stray lights affected dark I–V measurement

**Fig. 2e** is not fully representative of the intrinsic dark current of the device and is partly an artefact of the initial measurement conditions.

All I–V measurements in **Fig. 2e** were performed in a nominally light-tight enclosure; however, subsequent tests revealed that the measured “dark” current is likely overestimated due to residual background illumination. Possible sources include imperfect light sealing of the enclosure (e.g. small gaps at feedthroughs or door seams), scattered light from the 1550 nm beam path, and visible-wavelength indicator LEDs on the laser source and auxiliary equipment.

We confirmed this observation by re-measuring I–V characteristics (the same sample number 12) by placing the sample in a grounded Faraday cage to further suppress both residual optical background and electromagnetic interference, and the resulting dark current is obviously lower as compared in **Fig.S3**.

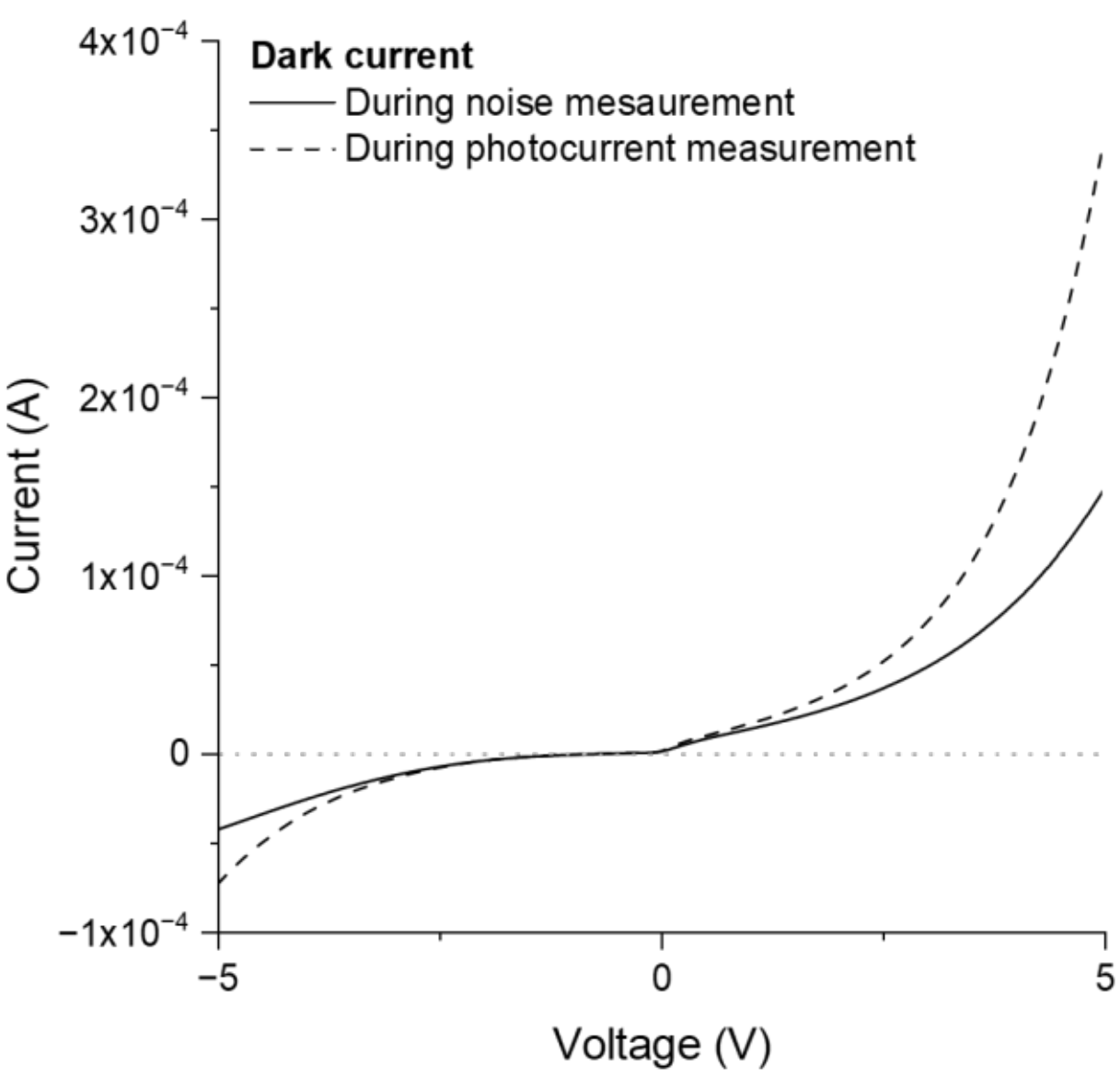

**Fig. S3 |** Dark current comparison under photocurrent (in a nominal dark box) and noise measurement (in a Faraday cage) setups.

The physical reason for this difference is the hyperdoped-Si detectors exhibit significantly higher responsivity at above-bandgap wavelengths than at 1550 nm (**Fig.S2**). As a result, even weak stray visible lights can generate a measurable photocurrent. This stray-light-induced photocurrent adds to the genuine dark current and leads to an overestimation of the “dark” curve.

Therefore, subtracting photocurrent solely from these DC sweeps using a standard SMU has fundamental limitations for weak sub-bandgap signals, especially when the testing environment contains weak stray lights with above-bandgap wavelengths. Specifically, there are two limitations:

(1) stray lights add to the dark current and therefore an underestimation of detectivity in general. (2) If the dark current is much higher than the photocurrent especially with higher bias voltage, the SMU is forced into a higher range which limits the separation of small signal for high DC current.

## Supplementary Text 5. Calculation of linear dynamic range (LDR)

At a forward bias of +5 V, we characterized the photocurrent as a function of incident 1550 nm power over nearly three orders of magnitude, from 4.28 µW to 4.87 mW. The photocurrent $I_{\mathrm{ph}}$ was obtained by subtracting the dark current from the total current at each power level.

A power-law fit of the form[24]

$$I_{\mathrm{ph}} \propto P^{\alpha}$$

over the range 4.28 µW–4.01 mW yields $\alpha \approx 0.97$, indicating a nearly linear response across this interval.

Using the minimum ($I_{\mathrm{min}}$) and maximum ($I_{\mathrm{max}}$) photocurrents in the linear response interval, we extract the linear dynamic range[24] as

$$\mathrm{LDR} = 20\log_{10}\left(\frac{I_{\mathrm{max}}}{I_{\mathrm{min}}}\right) \approx 59.4\ \mathrm{dB}.$$

Note that, in this study, the lower bound $I_{\mathrm{min}}$ is limited by the available attenuator, so the extracted LDR reflects the measurement window and should be interpreted as an apparent LDR rather than the intrinsic device capability[24].

## Supplementary Text 6. Calculation of normalized photocurrent

For the frequency response, we define a normalized photocurrent to compare the relative response under different conditions.

The normalized photocurrent in decibels is given by[24]

$$I_{\mathrm{ph}}(\mathrm{dB}) = 20\log_{10}\left(\frac{I_{\mathrm{ph}}}{I_{\mathrm{ph,max}}}\right),$$

where $I_{\mathrm{ph}}$ is the measured photocurrent amplitude at the frequency or preamplifier setting of interest, and $I_{\mathrm{ph,max}}$ is the maximum photocurrent amplitude observed under reference conditions (typically at low modulation frequency and/or at a high-sensitivity setting where the response is not bandwidth-limited).

This normalization removes the absolute scaling of the signal and highlights the relative attenuation as a function of modulation frequency or preamplifier gain. This allows curves taken at different powers or sensitivities to be plotted on the same scale and compared directly.

## Supplementary Text 7. Calculation of empirical −3 dB cutoff frequency

The rise (and fall) time recorded from the oscilloscope are defined as the intervals required for the signal to change from 10% to 90% (and 90% to 10%, respectively) of its steady-state amplitude[24].

For a first-order, linear low-pass system, the −3 dB cutoff frequency $f_{\mathrm{c}}$ (the frequency at which the output electrical power drops to 50% of its low-frequency value) can be approximately related to the 10–90% rise time $t_{\mathrm{r}}$ of the step response via the empirical relation[25]

$$f_c \approx \frac{0.35}{t_r}.$$

We measured $t_r \approx 70\ \mu s$ at 0 V and $t_r \approx 166\ \mu s$ at +5 V (with similar fall times of 63 and 186 µs, respectively). Substituting these values yields

$$f_c(0V) \approx \frac{0.35}{70 \times 10^{-6}} \approx 5 \text{ kHz},$$
$$f_c(5V) \approx \frac{0.35}{166 \times 10^{-6}} \approx 2.1 \text{ kHz},$$

Although the illumination is square-wave modulated (so each period contains multiple harmonics), the rising and falling edges still approximate step excitations. The 10–90% edge times therefore provide a suitable empirical basis for estimating $f_c$, and the resulting values are in reasonable agreement with the cutoff frequencies obtained directly from the frequency-response measurements (**Fig. 3f**).

## Supplementary Text 8. Noise measurement and specific detectivity determination

This section describes how the current-noise spectra shown in the main text are extracted from time-domain oscilloscope data using FFT-based methods, as schematically shown in **Fig.S4**. The procedure follows standard practice in spectral noise analysis. Note that during noise measurement, the sample was placed in a dark Faraday cage without stray lights and I-V was also measured with a SMU for theoretical noise floor calculation.

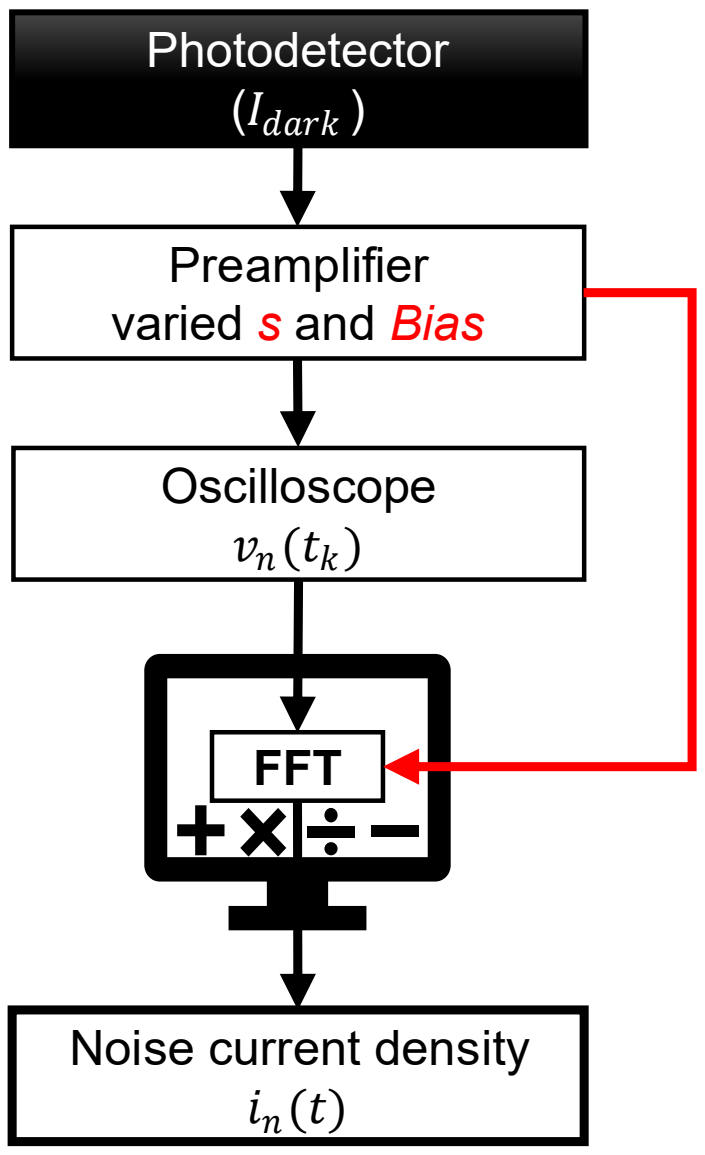

**Fig. S4** | Schematic illustrating the procedure used to convert oscilloscope voltage noise to input-referred current-noise density through Fast Fourier Transform (FFT) analysis.

The detector–preamplifier output was connected to a digital oscilloscope, which sampled the noise voltage $v_n(t_k)$ at 6.25 kS/s over a total acquisition time of 20 s, yielding a discrete time series with $t_k = k\Delta t$, where $\Delta t = 1/6.25$ kHz is the sampling interval and $k$ is an integer index.

The sampling rate is $F_s = 1/\Delta t$, and the corresponding Nyquist frequency, which sets the highest resolvable frequency, is $f_{\text{Nyq}} = F_s/2$. All spectra are therefore defined on the interval $0 \le f \le f_{\text{Nyq}}$. Before spectral analysis, we remove slow drifts that may otherwise contaminate the low-frequency noise. First, the global DC offset is subtracted by replacing $v(t_k)$ with $v'(t_k) = v(t_k) - \langle v \rangle$, where $\langle v \rangle$ is the time-averaged mean. Second, for each analysis segment used in the FFT, we detrend by subtracting a best-fit line, $v_s(n) \to v_s(n) - (an + b)$. This suppresses slow thermal or bias drift and prevents it from dominating the lowest Fourier bins.

To obtain a statistically reliable estimate of the spectrum, we employ Welch's method[26] rather than a single FFT of the entire record. The detrended time series is divided into $K$ overlapping segments of length $L$. For each segment we apply a window function, compute its discrete Fourier transform, form a one-sided power spectral density (PSD), and then average the PSDs over all segments.
If $K$ segments are averaged, the relative standard deviation of the PSD magnitude scales approximately as $\sigma_S/S \approx 1/\sqrt{K}$, so increasing the number of segments reduces the estimator variance. We use a Hann window,

$$w(n) = 0.5\left[1 - \cos\left(\frac{2\pi n}{L-1}\right)\right], n = 0, \dots, L-1,$$

and normalize the PSD by the mean-square window power,

$$U = \frac{1}{L}\sum_{n=0}^{L-1} w^2(n),$$

which ensures that the resulting spectrum has the correct units and correctly represents the noise power per unit bandwidth.

For each segment, the one-sided voltage PSD is computed as

$$S_{vv}(f_k) = \frac{2}{F_S\, L\, U} \mid X(f_k) \mid^2,$$

with units of $V^2$/Hz, where $X(f_k)$ is the discrete Fourier transform of the windowed segment. The corresponding voltage noise density (amplitude spectral density) is

$$v_n(f) = \sqrt{S_{vv}(f)},$$

with units of V/√Hz. The preamplifier is operated in transimpedance mode with a specified "sensitivity" $S$ in A/V, which corresponds to a nominal transimpedance $Z_t = 1/S$ (V/A). Assuming that the gain is approximately flat over the analysis band, the current-noise density is obtained from

$$i_n(f) = \frac{v_n(f)}{Z_t} = s\, v_n(f),$$

with units of A/√Hz. This is the quantity plotted and analysed in the main text.

The frequency resolution is set by the segment length. For a segment of $L$ points sampled at $F_s$, the frequency spacing is $\Delta f = F_S/L = 1/T_{\text{seg}}$, where $T_{\text{seg}} = L/F_S$ is the segment duration. In practice, the lowest usable frequency is limited both by the total record length and by detrending; we typically take $f_{\min} \approx \max(1/T_{\text{rec}}, 5\,\Delta f)$, where $T_{\text{rec}}$ is the total acquisition time, to avoid bins strongly affected by detrending and windowing. At the high-frequency end, we restrict our analysis to $f_{\max} \approx 0.8\, f_{\text{Nyq}}$, leaving margin below Nyquist to avoid aliasing and window-edge artefacts.

To characterize the noise over a finite frequency band $[f_1, f_2]$, we define a band-integrated current-noise density by integrating the squared spectral density $i_{\text{n}}(f)$ and normalizing by the bandwidth:

$$i_{\text{n,band}} = \sqrt{\frac{1}{f_2 - f_1}\int_{f_1}^{f_2} i_{\text{n}}^2(f)\, \text{d}f}.$$

This quantity has units of $\text{A Hz}^{-1/2}$ and represents the equivalent white-noise density that would produce the same total noise power over the band $[f_1, f_2]$.

In practice, the spectra are approximately flat (white) over the bands of interest, so we estimate $i_{\text{n,band}}$ by taking the average of $i_{\text{n}}(f)$ within the band and using the standard deviation as an uncertainty measure.

Using the sensitivity-dependent $i_{\text{n,band}}(S)$, together with the responsivity $R(S)$ obtained from the photocurrent measurements and the detector active area ($A = 0.49\ \text{cm}^2$) , we compute the specific detectivity

$$D^*(S) = \frac{R(S)\sqrt{A}}{i_{\text{n,band}}(S)}$$

for both high and low incident powers at 1550 nm. In this formulation, $D^*$ is expressed in Jones ($\text{cm Hz}^{1/2}\ \text{W}^{-1}$).

## Supplementary Text 9. Calculation of specific detectivity from I–V measurement and limitation

To determine the detector noise floor, we calculate shot noise and thermal (Johnson-Nyquist) noise from dark I–V measurement, and combine these contributions in quadrature:

$$i_{\mathrm{n}} = \sqrt{i_{\mathrm{shot}}^2 + i_{\mathrm{th}}^2},$$

with

$$i_{\mathrm{shot}} = \sqrt{2qI_{\mathrm{dark}}}, i_{\mathrm{thermal}} = \sqrt{\frac{4k_{\mathrm{B}}T}{R_{\mathrm{d}}}}.$$

Here, $q$ is the elementary charge, $I_{\mathrm{dark}}$ is the dark current at the bias of interest, $k_{\mathrm{B}}$ is Boltzmann's constant, $T$ is the absolute temperature, and $R_{\mathrm{d}}$ is the differential resistance extracted from the slope of the dark I–V curve at that bias:

$$R_{\mathrm{d}}(V) = \left|\frac{\mathrm{d}I_{\mathrm{dark}}}{\mathrm{d}V}\right|^{-1}.$$

We note that at lower modulation frequencies, additional noise sources such as 1/*f* noise and trap-related generation–recombination noise[27] may further increase the low-frequency noise. However, in our measurements, detectivity is evaluated for applications involving modulated optical signals in the kHz range, with shot noise becoming dominant at large bias and thus high dark current. Therefore, our approach results in a lower limit of the noise floor when the device operates at white noise plateau.

## SUPPLEMENTARY REFERENCE